\documentclass[a4paper]{jpconf}
\usepackage{graphicx}
\begin{document}

\title{Systematic errors in transport calculations of shear viscosity using the Green-Kubo formalism}

\author{J B Rose$^1$, J M Torres-Rincon$^1$, 
D Oliinychenko$^1$, A Sch\"afer$^1$ and H~Petersen$^{1,2,3}$}

\address{$^1$Frankfurt Institute for Advanced Studies, Ruth-Moufang-Strasse 1, 60438
Frankfurt am Main, Germany}
\address{$^2$Institute for Theoretical Physics, Goethe University,
Max-von-Laue-Strasse 1, 60438 Frankfurt am Main, Germany}
\address{$^3$GSI Helmholtzzentrum f\"ur Schwerionenforschung, Planckstr. 1, 64291
Darmstadt, Germany}

\ead{rose@fias.uni-frankfurt.de}

\begin{abstract}
The purpose of this study is to provide a reproducible framework in the use of the Green-Kubo formalism to extract transport coefficients. More specifically, in the case of shear viscosity, we investigate the limitations and technical details of fitting the auto-correlation function to a decaying exponential. This fitting procedure is found to be applicable for systems interacting both through constant and energy-dependent cross-sections, although this is only true for sufficiently dilute systems in the latter case. We find that the optimal fit technique consists in simultaneously fixing the intercept of the correlation function and use a fitting interval constrained by the relative error on the correlation function. The formalism is then applied to the full hadron gas, for which we obtain the shear viscosity to entropy ratio.
\end{abstract}

\section{Introduction}

There is currently great interest in the extraction of transport coefficients of hot and dense matter in the field of heavy ion physics. At temperatures lower than 170 MeV, in the hadron gas phase, previous studies of shear viscosity have however proven to be inconsistent with each other \cite{Muronga:2003tb,Demir:2008tr,Romatschke:2014gna,Pratt:2016elw}. In \cite{visco_paper}, more extensive results are discussed and the discrepancy is resolved by identifying the effect of resonance lifetimes on relaxation dynamics. Within these proceedings, a study of the systematic errors occurring in the application of the Green-Kubo formalism ~\cite{kubo,zubarev} is shown, with the
specific aim of providing a reproducible framework for the extraction of shear viscosity. To this end, we will use a newly developed transport model, Simulating Many Accelerated Strongly-interacting Hadrons, or SMASH~\cite{Weil:2016zrk}, in which infinite matter simulations for different
chemical compositions are carried out.

\section{Green-Kubo formalism}

The Green-Kubo formalism is a  method to extract transport coefficients from fluctuations of given currents around a state of equilibrium in a given system. In the case of shear viscosity,
\begin{equation}
  \eta = \frac{V}{T} \int_0^\infty dt \ C^{xy} (t),
  \label{kubo_shear_def}
\end{equation}
where $V$ is the volume, $T$ the temperature and $t$ the time. The auto-correlation function $C^{xy} (t)$ is given by
\begin{equation}
  C^{xy} (t) = \langle T^{xy} (0) T^{xy} (t) \rangle_{eq}
                  = \lim_{N \rightarrow \infty} \frac{1}{N} \ \sum_{s=0}^{N} T^{xy}(s) \ T^{xy} (s+t),
  \label{correlator}
\end{equation}
where $T^{xy}$ is an off-diagonal component of the space-averaged energy-momentum tensor and $N$ the number of timesteps used in taking the average. In all our calculations, this is set to $N=5000$ for
$C^{xy} (0)$, with a timestep length of $\delta t = 0.05$ fm, so the total interval used to compute the correlation function is 250 fm. The energy-momentum tensor is computed as
\begin{equation}
  T^{\mu\nu} = \frac{1}{V} \sum_{i=1}^{N_{part}} \frac{p^\mu_i p^\nu_i}{p^0_i} \ ,
  \label{tmunu_discrete}
\end{equation}
where $N_{part}$ is the total number of particles and $p^{\mu}$ a component of the momentum
4-vector of a given particle.

\begin{figure}[h]
\begin{center}
\begin{minipage}[t]{75mm}
\includegraphics[width=75mm]{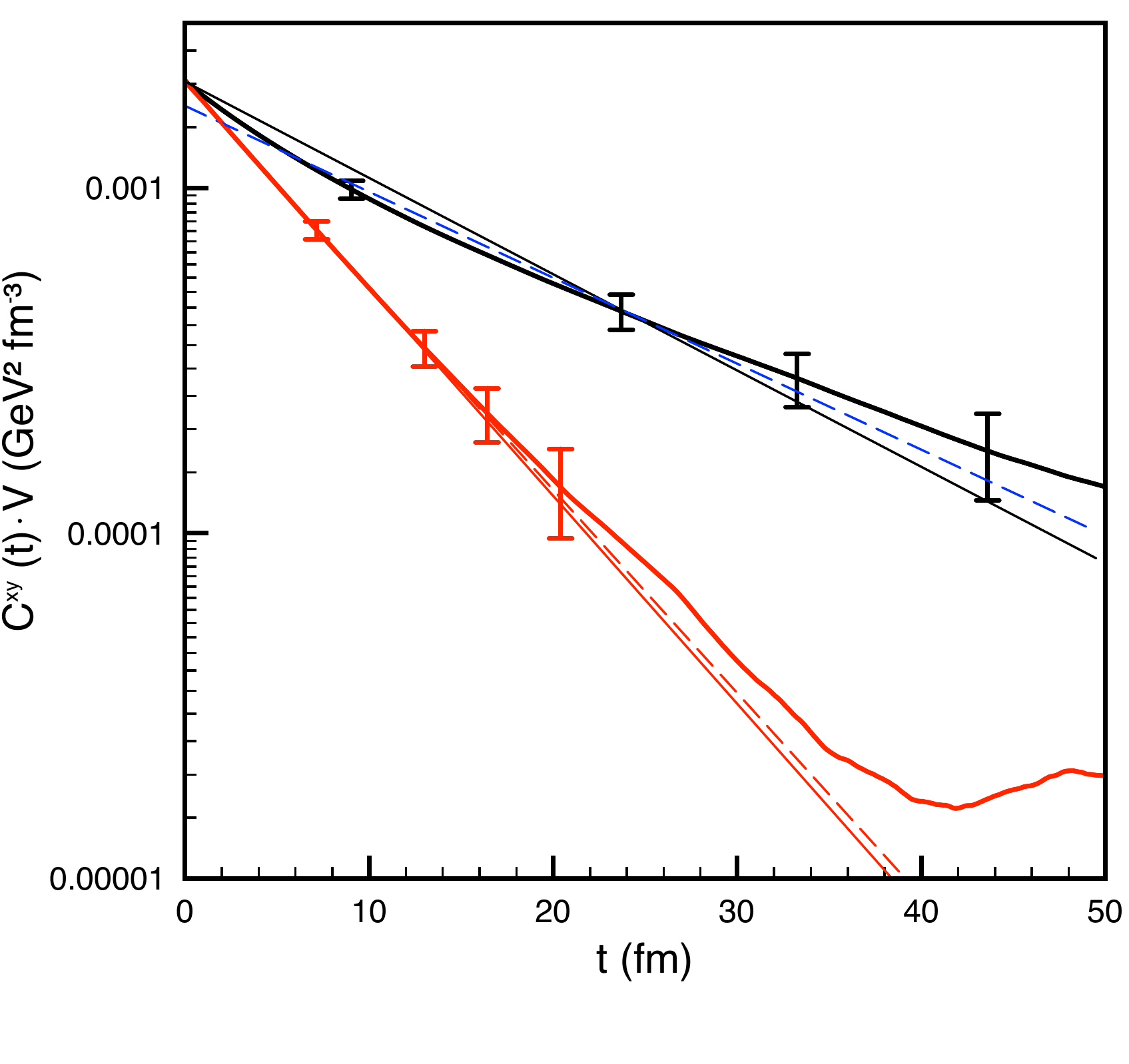}
\caption{\label{correl_evol}Average volume-independent correlation function as a function of time for a system of particles with constant cross-section (red, bottom set of curves) and energy-dependent pion cross-section (black, top set of curves). Thinner lines are exponential fits, respectively with (solid) and without (dashed) a fixed intercept.}
\end{minipage}\hspace{2pc}%
\begin{minipage}[t]{75mm}
\includegraphics[width=75mm]{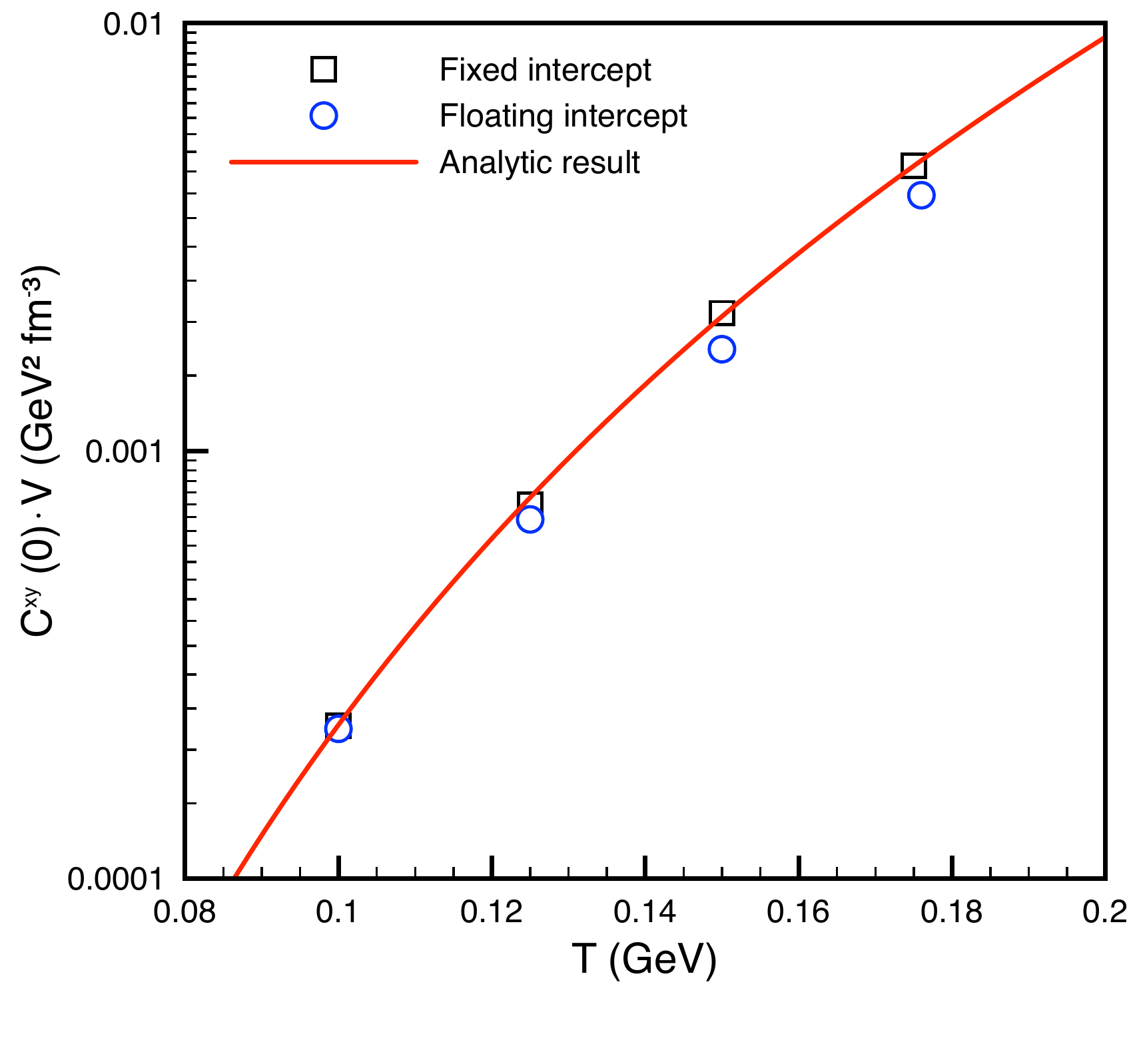}
\caption{\label{correl_zero}Volume-independent initial value of the correlation function as a function of temperature for a system of particles interacting via the energy-dependent pion cross-section. Statistical error bars are smaller than the symbol size.}
\end{minipage}
\end{center}
\end{figure}

\section{Correlation functions and shear viscosity of simple systems}

As one can see from Eq. (\ref{kubo_shear_def}), the calculation of the viscosity requires integrating $C^{xy} $ from
zero to infinity. Numerically, it is challenging to take the limit of $N \to \infty$ in Eq.~(\ref{correlator}).
Consequently, the relative error of any numerical computation of the correlation function necessarily increases 
rather quickly with $t$ and eventually reaches a state of pure noise, as one can already see in the thick red curve 
of Fig. \ref{correl_evol}. To circumvent this limitation, some assumption is made about the analytical 
shape of the correlation function.

It is generally assumed~\cite{Demir:2008tr,Wesp:2011yy,Plumari:2012ep} that for dilute systems, it takes the form
of a decaying exponential,
\begin{equation}
  C^{xy} (t) = C^{xy} (0) \ e^{-\frac{t}{\tau}} \ ,
  \label{correl_ansatz}
\end{equation}
where $\tau$ is the relaxation time of the system. For the shear viscosity, it follows that
\begin{equation}
  \eta = \frac{C^{xy} (0) V \tau}{T} \ .
  \label{final_shear_eq}
\end{equation}

The thick lines of Fig. \ref{correl_evol} correspond to the average correlation functions of two different test systems at $T=150$ MeV. In the first (red), massive particles ($m=138$ MeV) interact through a constant isotropic cross-section of $\sigma=20$ mb, whereas in the second (black), the same particles interact through an isotropic energy-dependent cross-section corresponding to the $\rho$ resonance. In all cases here and further on, the correlation function is obtained by averaging 1000 simulations of the same system.
The behavior of the correlation function of these two systems is definitely not the
same. While in the case of the constant cross-section interaction, the behavior is quite close to the previously 
mentioned decaying exponential, we observe a slight deviation from it in the energy-dependent case. This phenomenon is more apparent at high temperatures and densities; the deviation of the shown curve is thus more pronounced than most others used in the rest of this paper, for the purpose of illustration.

\begin{figure}[h]
\begin{center}
\begin{minipage}[t]{75mm}
\includegraphics[width=75mm]{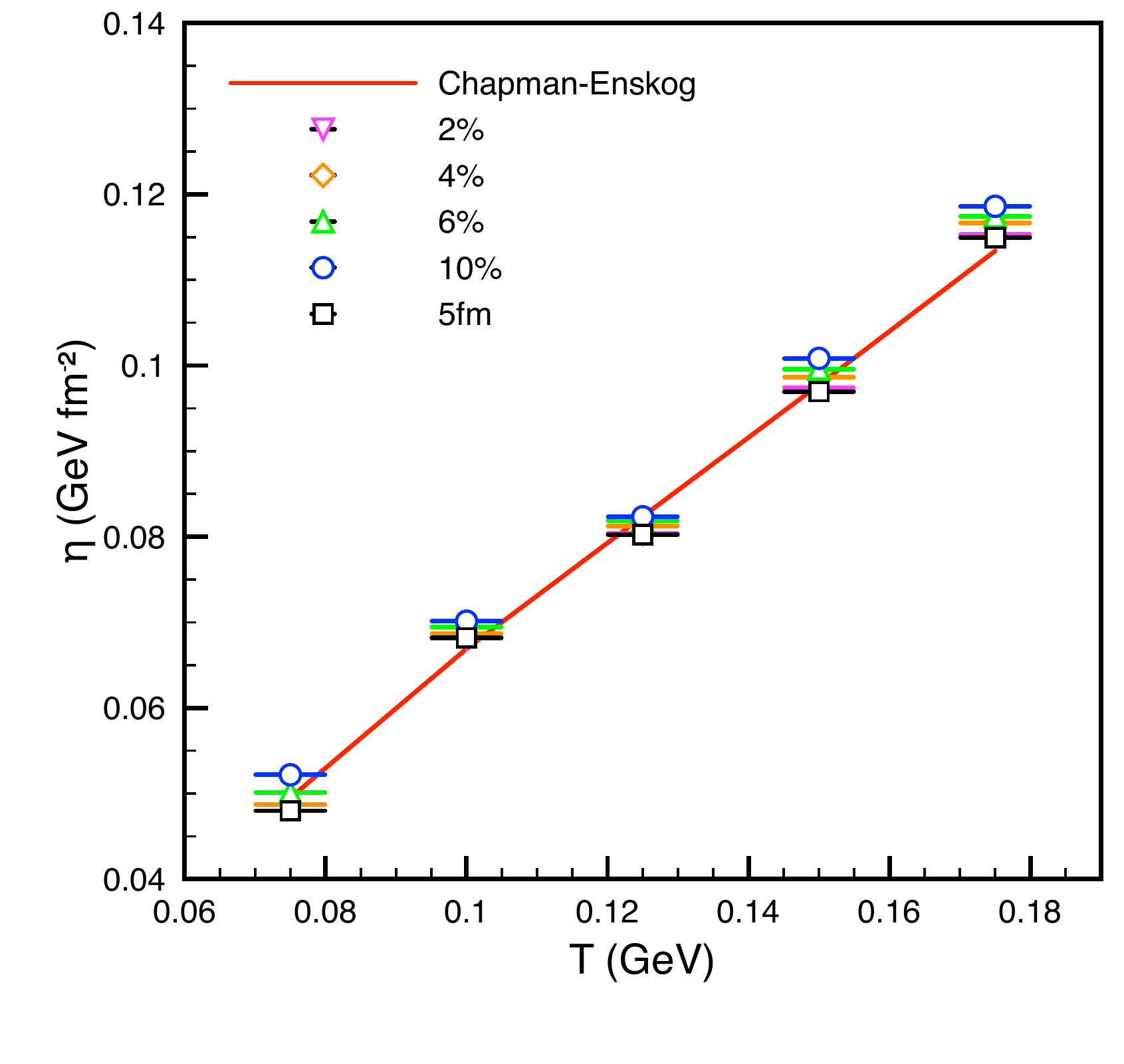}
\caption{\label{fit_distances_constant}Shear viscosity of a system of particles interacting via constant cross-section for various fitting schemes, compared to a semianalytical Chapman-Enskog calculation. ($\sigma=20$ mb, $m=138$ MeV)}
\end{minipage}\hspace{2pc}%
\begin{minipage}[t]{75mm}
\includegraphics[width=75mm]{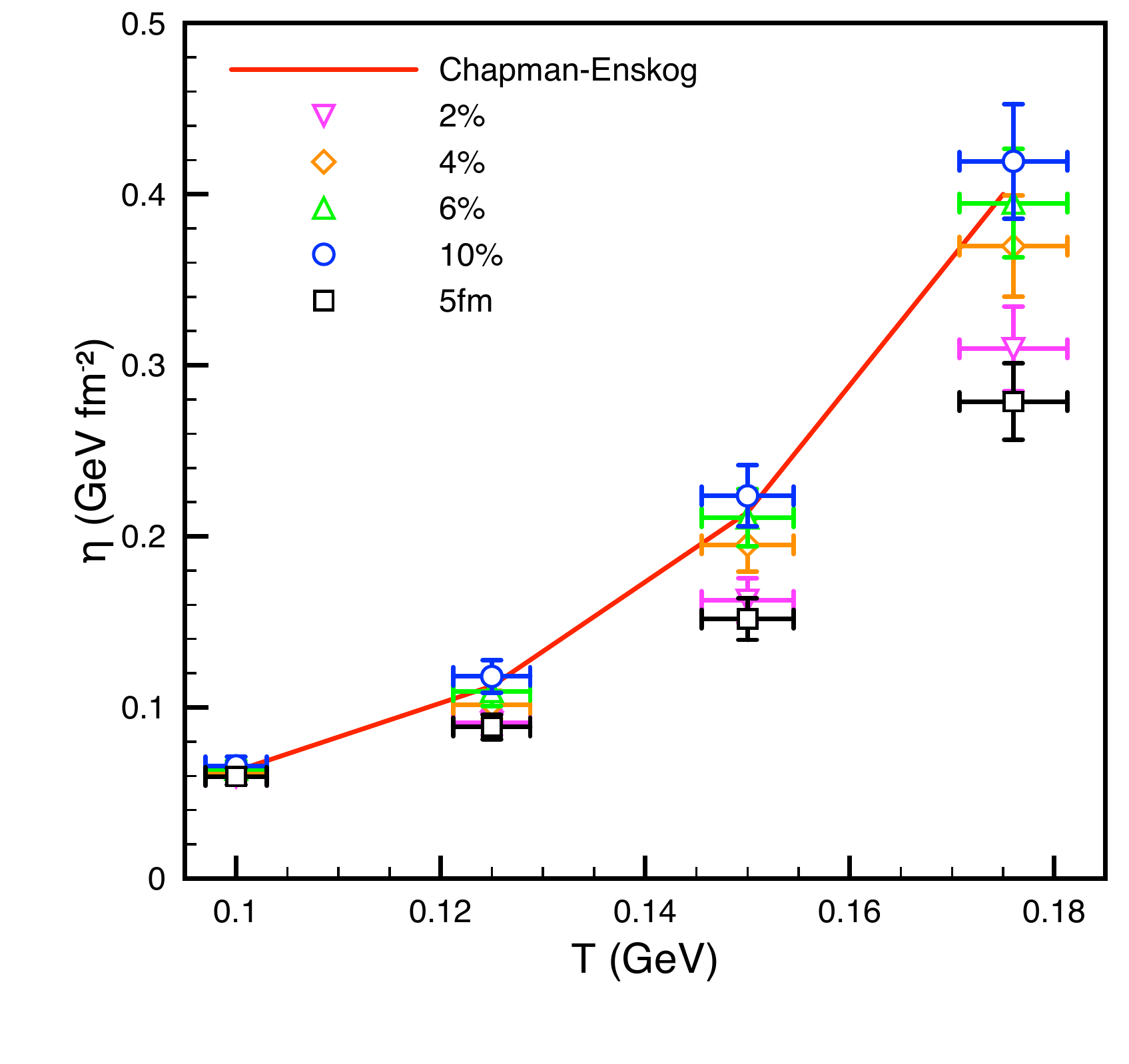}
\caption{\label{fit_distances_rho}Shear viscosity of a system of particles ($m=138$ MeV) interacting via the energy-dependent pion cross-section for various fitting schemes, compared to a semianalytical Chapman-Enskog calculation.}
\end{minipage}
\end{center}
\end{figure}

For $N_{spec}$ stable particles, the initial value of the correlation function is given by taking the continuum limit of
\begin{equation}
  C^{xy} (0) = \langle \sum_i \frac{(p_i^x)^2
  (p_i^y)^2}{V^2(p_i^0)^2} \rangle \rightarrow
  \sum_{a=1}^{N_{spec}} \frac{g_a z_a}{30 \pi^2 V} \int_0^\infty dp \
  \frac{p^6}{m_a^2+p^2} \exp \left( - \frac{\sqrt{m_a^2+p^2}}{T} \right) \ ,
  \label{cxy0}
\end{equation}
where $z_a=\exp(\mu_a/T)$ is the fugacity of the species $a$ and $g_a$ its degeneracy factor. 
Since $C^{xy} (0)$ is known analytically, it is reasonable to fix it in the fit of the correlation function. Indeed, Fig. 
\ref{correl_zero} shows that a fit using a floating intercept systematically undershoots the analytical value of $C^{xy} (0)$ at higher temperatures. Even though such a fit might lead to a better fit of the original curve, a precise value for the initial value is required in our final viscosity calculation (Eq. (\ref{final_shear_eq})). Therefore, $C^{xy} (0)$ needs to be fixed.

Let us now turn our attention to the question of how many points to consider when fitting. We tried two different schemes to this effect. The first one is to consider a fixed interval from $t=0$ to $t=5$ fm; this interval is small enough that it always very closely fits the earliest part of the curve. The second scheme takes into account that the growth rate of the relative error of the correlation is much larger at higher temperatures, which is done 
using a fitting interval for which the cutoff depends on this relative error. This is illustrated by the four points which have error bars on both thick curves of figure \ref{correl_evol}: they correspond in each case to the point where the relative statistical error reaches 2\%, 4\%, 6\% and 10\%. The plotted fits correspond here in all cases to using all points up to 6\% relative error.

In order to determine which of these cutoffs to use, we now compare the final shear viscosity yielded by each to an analytical calculation of the shear viscosity of the previous two simple systems using the Chapman-Enskog approximation to solve the Boltzmann equation~\cite{Torres-Rincon:2012sda}. Fig. \ref{fit_distances_constant} shows the effect of varying the cutoff on shear viscosity in the case of a system interacting through constant cross sections, whereas Fig. \ref{fit_distances_rho} shows the same for the case of energy-dependent interactions. In the second case, resonance lifetimes have been decreased to zero in SMASH, so that the comparison is carried out between comparable systems using point-like 2-to-2 interactions. All calculations use an isotropic cross-section.

As expected from looking at the correlation function on Fig. \ref{correl_evol}, the final effect of varying the cutoff is 
rather limited when cross-sections are isotropic. All proposed cutoff schemes and values appear to describe well 
the analytical calculation, with the largest deviations remaining under 8\% in all cases. This value is then taken as 
a systematic error on the method for all further calculations (note that statistical error bars are smaller than the 
symbol size). If one now looks at Fig. \ref{fit_distances_rho}, where cross-sections are energy dependent, the 
picture is different. It is here very clear there are cases where the sensitivity to the cutoff is large. Cutoffs at 
4\%, 6\% and 10\% relative error manage to fit the Chapman-Enskog calculation within systematic errors. While it 
appears that in this regime it is still possible to use exponential fits, one should keep these deviations in mind when 
using this ansatz, and possibly look into different ones if deviations become larger.

\section{Shear viscosity of a hadron gas}

\begin{figure}[h]
\begin{center}
\begin{minipage}[t]{75mm}
\includegraphics[width=75mm]{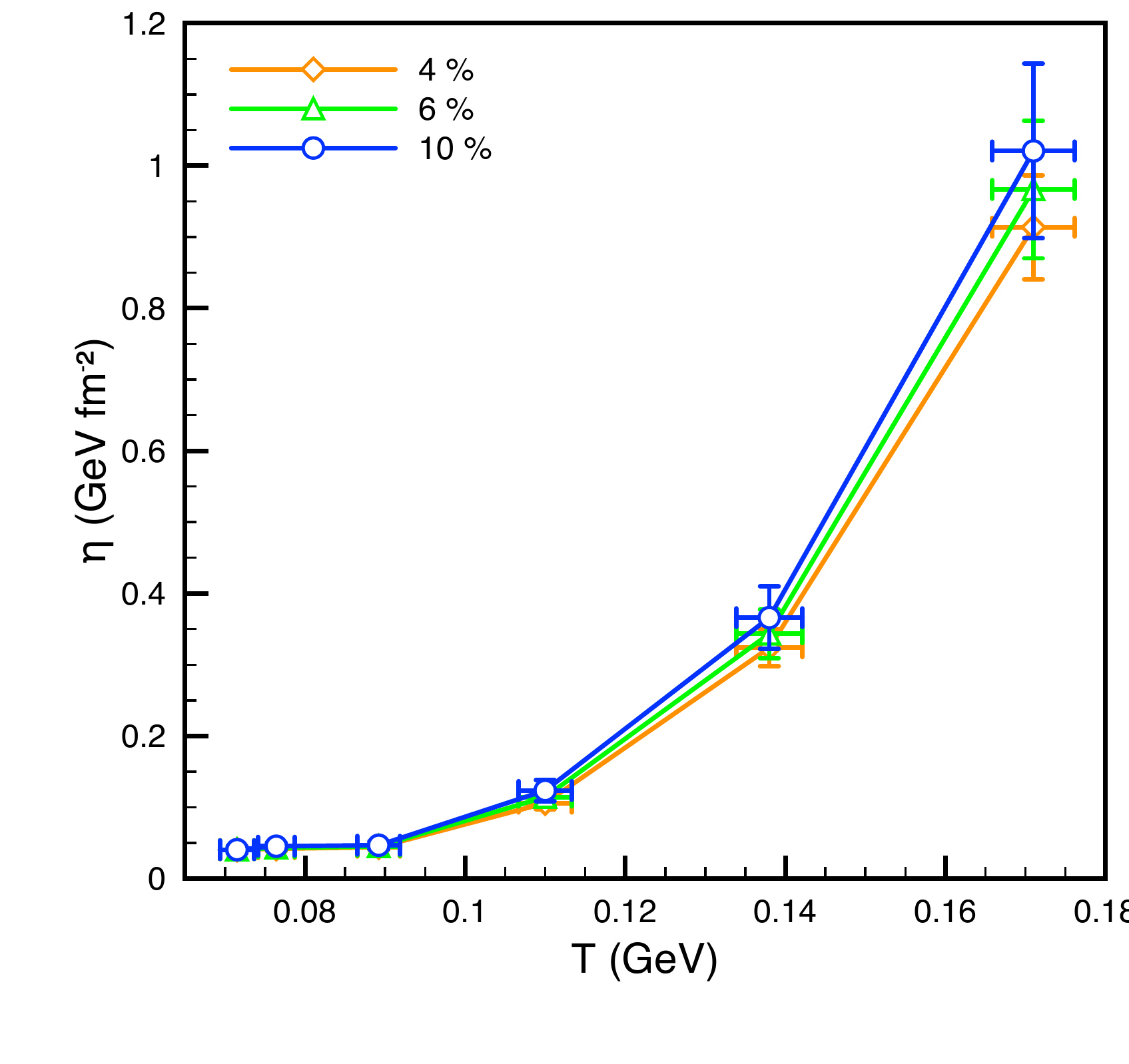}
\caption{\label{shear_hadron}Shear viscosity of a full hadron gas as a function of temperature, for various fitting schemes, at $\mu_B=0$ MeV.}
\end{minipage}\hspace{2pc}%
\begin{minipage}[t]{75mm}
\includegraphics[width=75mm]{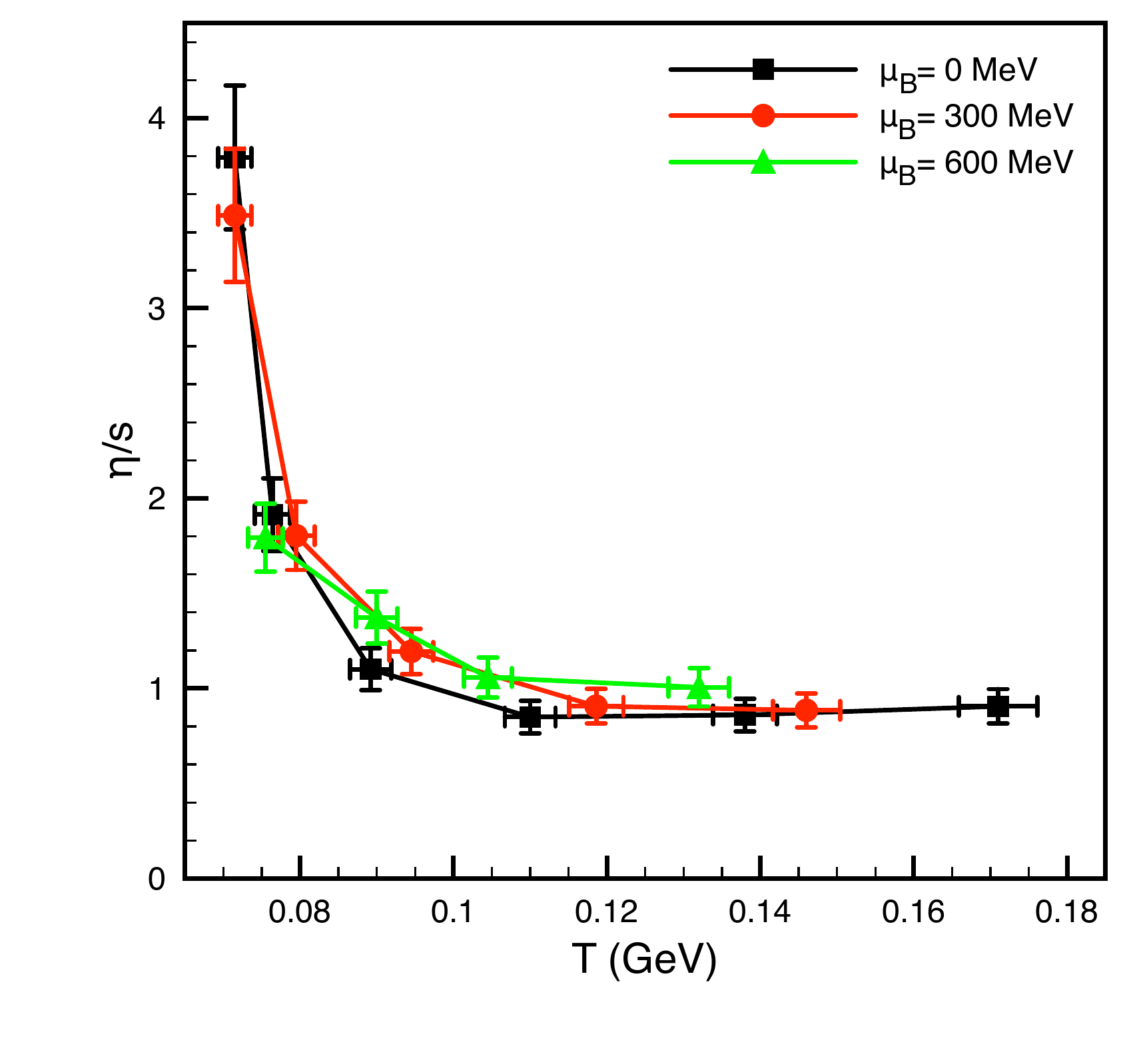}
\caption{\label{full_viscosity}Shear viscosity over entropy density of a full hadron gas, for various baryochemical potentials.}
\end{minipage}
\end{center}
\end{figure}

As a final result, the viscosity of the full hadron gas is calculated within SMASH (Fig. \ref{shear_hadron}) at $\mu_B=0$ MeV. Once again, the computation is performed using the three previously mentioned cutoffs that were accurately describing analytical calculations. In the case of the full hadron gas, however, such a comparison is not feasible, as no analytical equivalent currently exists. We obtain in this regard results which are very similar to the previous case: although there is some variation in the actual value of the shear viscosity, the different cutoffs produce results which are consistent with each other within systematic errors. We thus conclude from this technical study that it is sufficient to use a cutoff corresponding to a relative error of 6\% when fitting the correlation function. All curves being mutually consistent, this corresponds to the middle point and reproduces best the available analytical calculations.

Fig. \ref{full_viscosity} shows the final result of this procedure: the ratio of shear viscosity to entropy density. Here the entropy density is calculated according to the Gibbs formula $s = \frac{\epsilon + p - n_B \mu_B}{T}$, where $\epsilon$ is the energy density, $p$ the pressure, and $n_B$ and $\mu_B$ respectively the baryonic net density and chemical potential. We observe that the profile of this quantity is decreasing, which is the expected behavior when approaching a phase transition, and eventually reaches a plateau. This behavior is consistent with the results in Ref.~\cite{Demir:2008tr}. We furthermore observe a very mild dependence on baryonic chemical potential, although all results are here within systematic error of each other. This could then also be consistent to an independence on chemical potential up to these values, as previously found in Ref.~\cite{Itakura:2007mx}.

\ack
This work was made possible thanks to funding from the Helmholtz Young Investigator Group VH-NG-822 from the Helmholtz Association and GSI, and supported by the Helmholtz International Center for the Facility for Antiproton and Ion Research (HIC for FAIR) within the framework of the Landes-Offensive zur Entwicklung Wissenschaftlich-Oekonomischer Exzellenz (LOEWE) program launched by the State of Hesse. Computing services were provided by the Center for Scientific Computing (CSC) of the Goethe University Frankfurt.

\end{document}